\documentclass{aa}
\usepackage{graphicx}

\begin{document}

\sloppypar

   \title{Intensity of the cosmic X-ray backgound from HEAO1/A2 experiment}

   \author{M. Revnivtsev\inst{1,2}, M. Gilfanov \inst{1,2}, K.Jahoda \inst{3}, R. Sunyaev \inst{1,2}}

   \offprints{mikej@mpa-garching.mpg.de}

   \institute{Max-Planck-Institute f\"ur Astrophysik,
              Karl-Schwarzschild-Str. 1, D-85740 Garching bei M\"unchen,
              Germany,
        \and   
              Space Research Institute, Russian Academy of Sciences,
              Profsoyuznaya 84/32, 117810 Moscow, Russia
        \and 
                Laboratory for High Energy Astrophysics, Code 662, Goddard Space Flight Center, Greenbelt, MD 20771, USA
            }
  \date{}

        \authorrunning{Revnivtsev et al.}
        \titlerunning{}
 
   \abstract{We reanalyze data of HEAO1/A2 -- Cosmic X-ray Experiment -- 
in order to repeat the measurements of the cosmic X-ray background (CXB) 
intensity and 
accurately compare this value with other measurements of the CXB. 
We used the data of
MED, HED1 and HED3 detectors in scan mode, that allowed us to measure
effective solid angles and effective areas of detectors self consistently, 
in the same mode as the CXB intensity was measured. We obtained the average 
value of the CXB intensity is $1.96\pm0.10 \times 10^{-11}$ erg s$^{-1}$ 
cm$^{-2}$ deg$^{-2}$ in the energy band 2-10 keV, or 
$9.7\pm0.5$ phot s$^{-1}$ cm$^{-2}$ at 1 keV assuming the power law 
spectral shape with photon index $\Gamma=1.4$ in this energy band. 
We compare the obtained measurements with 
those obtained by different instruments over last decades.
   }

   \maketitle

%

\section{Introduction}

Emission of extragalactic X-ray sources, discovered in 1962 
(\cite{giacconi62}) as the cosmic X-ray background (CXB) still 
remain one of the most interesting topic
of X-ray astronomy and observational cosmology. 
Over last decades it was shown that the cosmic 
X-ray background consists of emission of a large number of point sources 
(see e.g. \cite{giacconi02}), mostly
 active galactic nuclei (AGNs). The focusing telescopes like EINSTEIN, 
ROSAT, CHANDRA and XMM
have resolved most of the CXB into separate point-like objects. 
In the view of such a progress in this field a special attention is paid now 
to the accurate measurements of the CXB intensity value. During last decades
many different instruments measured the intensity of the CXB and
still there are some discrepancies between the obtained values (see e.g. 
\cite{moretti03,mikej03} and references therein).

Likely the most important information about the shape and the average amplitude
of the CXB in X-ray energy range ($\sim$2--60 keV) is still based on the
measurements of HEAO1 observatory (1977-1979). 
The instrument Cosmic X-ray Experiment (also known as 
A-2 experiment) aboard this observatory was specially designed for 
accurate measurements of the CXB. The key
feature of this instrument was the ability to distinguish between the
internal instrumental background and the cosmic X-ray background (see e.g. 
\cite{rothschild79}, \cite{marshall80}, \cite{boldt87}).
The HEAO1 observatory spent a significant fraction of its lifetime
scanning the whole sky and allowed to construct the all sky survey
(e.g. \cite{piccinotti82,wood84,levine84}) and 
to measure the cosmic X-ray background over very wide sky solid angle
(e.g. \cite{marshall80,boldt87,gruber99}).

One of the main difficulties in the comparison of CXB results of different 
observatories is the accuracy of their cross-calibrations.
Proportional counters with collimators, which gave a lot of information 
about the CXB, are relatively easy to crosscalibrate if they work in 
similar energy bands. Ability to observe the Crab nebula, which
is now considered almost perfectly stable celestial source, 
allows one to use this source for straightforward crosscalibration of 
the instruments. Unfortunately, published information about the parameters
of HEAO1/A2 detectors is not sufficient for accurate crosscalibration of
its results with results obtained by modern satellites.

In this paper we reanalyze data of HEAO1/A2 experiment and obtain
the intensity of the CXB and important instrumental parameters of HEAO1/A2 
which allowed us to relatively accurately compare the results of HEAO1/A2 with 
those of other instruments.

\section{HEAO1/A-2 instrument}

Detailed description of the Cosmic X-ray Experiment (A2) aboard HEAO1 
observatory can be found in Rothschild et al. (1979). Here we only briefly
describe general features of the instrument. 

The A2 experiment (Cosmic X-ray Experiment) of HEAO1 observatory 
consisted of three sets of different types of detectors. All detectors 
were proportional counters with different filling
gas. Low Energy Detectors (LED) worked in the energy band 0.15--3 
keV, Medium Energy Detector (MED) had the effective energy band 1.5-20 keV 
and High Energy Detectors (HED) had the energy band 2--60 keV.

The largest advantage of the A2 was the ability to separate the internal 
instrumental background from the cosmic X-ray background. This was achieved 
by a special design
of detectors. All 6 detectors of the A2 were proportional counters with 
detective layers of anodes. 
Half of anodes were illuminated through
$\sim 3^\circ\times3^\circ$ collimators, another half through
$\sim 1.5^\circ\times3^\circ$ (or $\sim 6^\circ\times3^\circ$) collimators. The flux of the cosmic X-ray
background measured by detectors rises with their
collimators solid angles. On the other hand the instrumental background in 
different parts of the detectors which were under different collimators
was the same by design (two types of collimators were interlaced with each
other and detecting anodes under them were intermixed, see more detailed
description in Rothschild et al. 1979). 
The total flux, detected by different halfs of 
the detectors (which see the sky through different collimators): 
$C_{\rm L}=C_{\rm bkg}+C_{\rm CXB, L}$, $C_{\rm S}=C_{\rm bkg}+C_{\rm CXB, S}$.
Here $C_{\rm bkg}$ -- count rate of the internal background, $C_{\rm CXB}$ --
count rate produced by the cosmic X-ray background in large (L) and small (S) 
field of view parts of the detector. Keeping in mind that the internal 
background is the same for L and S parts of the detector, the 
the CXB flux detected by the ``small'' half of the detector ($C_{\rm CXB, S}$)
 can be calculated from the simple formula:

\begin{equation}
C_{\rm CXB, S}=(C_{\rm L}-C_{\rm S}) \left({{A_{\rm L}\Omega_{\rm L} \over{A_{\rm S}\Omega_{\rm S}}}-1} \right)^{-1}  \,\,\,\ \rm cnts/s
\end{equation}

in which  $A\Omega$ -- the production of effective area and solid angle
of L and S parts of the detector.

\section{Data analysis}
One of the main goal of HEAO1/A2 experiment was to measure the intensity
of the CXB averaged over large sky solid angle. For this purpose the satellite
rotated around the Sun-pointed axis (33 minutes period) 
which was gradually stepped every 12 hours by $\sim 0.5^\circ$ in order 
to remain pointed at the Sun.
The measurement of the CXB was based on formula (1) (see \cite{marshall80}).

During the scanning mode the HEAO1/A2 covered the whole sky 
(\cite{piccinotti82}) including the Crab nebula. This gives us an
opportunity to use the measurements of the Crab nebula for accurate
crosscalibration with other collimated instruments. Assuming the same
photon flux of the Crab nebula for all instruments we can rescale their
effective area (if needed), and using the scans of the HEAO1/A2 detectors 
over the source we can determine the collimators effective solid angles.

For our analysis we have used the HEAO1/A2 database in Goddard Space Flight 
Center (ftp://heasarc.gsfc.nasa.gov/FTP/heao1/data/a2/xrate\_fits).
The database provides the count rate measurement of all A2 detectors 
every 1.28 sec.
The count rate measurements are presented in the form of discovery scalers --
count rates of detectors in a certain energy bands
(for more detailed description see \cite{marshall83}, \cite{allen94})
Definition of some discovery scalers changed throughout the mission. Therefore
for our purpose we used only those scalers which have not been changed --
the total count rate of different parts of different anode layers of the 
detectors -- 1L, 1R, 2L and 2R. Here L and R denotes so called ''left'' and 
''right'' parts of the detectors, placed under different size collimators
(see Rothschild et al. 1979). 

We used only data of detectors MED, HED1 and HED3 due to the following reasons.
As we are interested in the hard X-ray background ($>$ 2 keV) we do not 
consider here LED detectors which worked in 0.15-3 keV energy band. 
We also do not consider data of detector 
HED2 which lacks the particle veto layer. 

The data were selected using the following criteria:
\begin{enumerate}
\item We analyzed only data obtained during scan mode. For HED3 detector
 we used only data before 304 day of the mission, for which period we have the 
response matrices of two separate layers of the the detector.
\item Data is ``clean'' -- detectors fields of view exclude the Earth plus 
100 km atmosphere, high voltage is on and stable,  
the calibrations rods  are outside MED field of view
\item Electron contaminations is not important 
\end{enumerate}

For subsequent analysis of the obtained count rate values we used
response matrices of A2, provided by the HEASARC archive 
(http://heasarc.gsfc.nasa.gov/FTP/heao1/data/a2/responses/).

\begin{figure}
\includegraphics[width=\columnwidth]{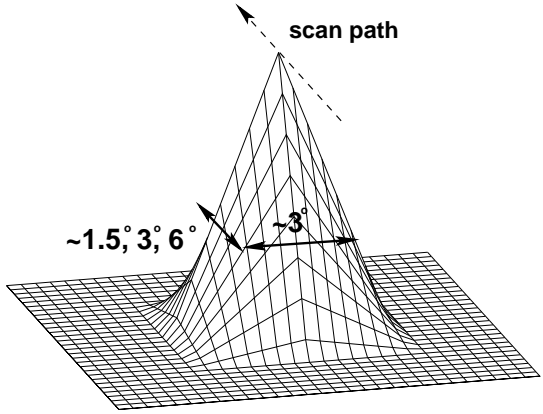}
\caption{Dependence of the effective area of HEAO1 detectors on the offset position
 of a source in the field of view.  
In the direction along the scan path 
collimators had FWHM $\sim 1.5 ^\circ$, $\sim 3^\circ$ and $\sim 6^\circ$ size.
Perpendicular to the scan path the size of all collimators had FWHM $\sim 3^\circ$.
 \label{collimator}}
\end{figure}

For determination of the collimators solid angles and effective areas
of the A2 detectors we used data of scans over the Crab nebula.
The A2 collimators are made of rectangular cross section tubes therefore
providing roughly linear dependence of the effective area of a detector
on the source offset within the field of view. 

\begin{equation}
C=CR \,\,\,[1-{\phi\over{\phi_0}}] [1-{\theta\over{\theta_0}}]
\end{equation}

here $C$ -- measured Crab nebula count rate at certain offset $\phi, \theta$; 
$CR$ - Crab nebula count rate observed by the detector at zero offset,
$\phi_0$ and $\theta_0$ -- size of the collimator field of view across
the scan plane (approximately $\sim 3^\circ$), and along the scan plane
($\sim 1.5^\circ, 3^\circ$, or $6^\circ$).
 Shape of the collimator off-axis 
response function is shown in Fig. \ref{collimator}.
Fluxes of the Crab nebula measured by every A2 detector at different 
offsets provide the shape of the collimator response function.
We have fitted the measured values by the described model (see formula (2))
using standard $\chi^2$ minimization technique. The quality of the fit in all cases
is good, the ratio of $\chi^2$ to the number of degrees of freedom
never exceeds 1.2.
Best fit parameters of the model
are presented in Table \ref{parameters}.

In order to convert the observed maximal Crab count rate ($CR$) into the 
effective area of the detectors (with the help of known response matrices) 
we assumed that the Crab spectrum is a power law ($dN/dE = A E^{-\Gamma}$) 
with $\Gamma=2.05$ and the normalization at 1 keV 
$A=10$ phot s$^{-1}$ cm$^{-2}$ keV$^{-1}$ (\cite{seward78,zombeck90}), similar 
to that used in the paper of Revnivtsev et al. (2003).
The energy flux of the Crab in this case equals to $2.39\times 10^{-8}$ 
erg s$^{-1}$ cm$^{-2}$ in the energy band 2-10 keV.  The obtained (measured)
Crab count rate ($CR$ in the formula (2)) depends on the response
matrix of the detector (which we took from HEASARC HEAO1 archive, see 
reference above)
and its effective area. 

 In the Table 
\ref{parameters} we present the best fit parameters of A2 MED, HED 1 and HED3 
detectors. 

Uncertainties of presented values have two main origins. The statistical
uncertainties are rather small. Typically they are not larger than $\sim$1\%.
The angular sizes of collimators and the effective solid angle 
can be determined with an accuracy $\sim$1\%. However, the values of 
effective areas of the detectors strongly depend on the assumed spectrum
of the Crab nebula and the accuracy of the used response matrices. 
For example, the difference around 0.05 in the photon index $\Gamma$ 
of the Crab nebula (leaving energy flux of the Crab in the energy band 2-10 keV 
unchanged) will change the count rate to energy flux conversion factor
by a 1-2\% percents for MED detectors and 2-3\% for HED detectors.
Difference in the assumed Crab nebula flux linearly translates into the
changes of effective area values.

\begin{table}
\caption{Parameters of detectors of A2, determined from the scanning observations.
\label{parameters}}
\begin{tabular}{l|c|c|c|c|c}
   &$\phi_0$, deg&$\theta_0$,deg&$\Omega$,deg$^2$&CR$^a$&$A_{\rm eff}$, cm$^2$\\
\hline
MED&&&&\\
layer1, left&2.88&2.93&8.44&602&405\\
layer1, right&2.92&1.42&4.15&539&362\\
layer2, left&2.86&2.93&8.38&269&366\\
layer2, right&2.90&1.43&4.09&243&331\\

\hline
HED1&&&&\\
layer1, left&3.15&2.84&8.95&559&335\\
layer1, right&3.15&5.87&18.49&564&338\\
layer2, left&3.16&2.88&9.10&55.6&331\\
layer2, right&3.13&5.86&18.34&55.4&324\\
\hline
HED3&&&&\\
layer1, left&3.10&2.96&9.17&607&354\\
layer1, right&3.09&1.48&4.56&577&336\\
layer2, left&3.05&2.89&8.81&58.5&340\\
layer2, right&3.04&1.49&4.53&53.7&313\\
\hline
\end{tabular}
\begin{list}{}
\item $^a$ -- Crab count rate in the detectors, cnts/s
\item -- All effective areas were calculated using the response 
matrices from HEAO1/A2 archive in GSFC. Namely m11n.rsp and m12n.rsp for two layers of
 MED, h111095c.rsp and h121095c.rsp for HED1, h31257c.rsp and h321095c.rsp for HED3 detectors.
\end{list}
\end{table}

\section{Results}

The main idea of measuring the intensity of the CXB
with the help of HEAO1/A2 instrument
is to use the fact that the flux of the CXB 
linearly scales with the solid angle of the detectors.
 Design of the A2 detectors allowed to exclude 
the instrumental
background with almost absolute accuracy (see details in \cite{rothschild79,boldt87}).
Measuring the average level of the difference between count rates of
large and small solid angle detectors
we can calculate the intensity of the CXB using formula (1). Coefficients
$A_{\rm L}\Omega_{\rm L}/(A_{\rm S}\Omega_{\rm S})$ can be 
calculated from Table 1.

For measuring the flux difference between the large and the small solid 
angle detectors we have used data from a part of the sky with galactic 
latitudes $|b|>20^\circ$
and also we excluded a region around very bright Galactic X-ray source 
Sco X-1 ($10^\circ$ around Sco X-1).
After this procedure some galactic sources still remain on the sky (e.g. Her X-1), 
however they contribute less than 1\% to the total sky flux.
Point-like extragalactic sources which can be detected from HEAO1/A2 survey 
(the flux is higher than $\sim 3\times 10^{-11}$ erg s$^{-1}$ cm$^{-2}$, see \cite{piccinotti82})
are part of the cosmic X-ray background, but in any case they do not 
contribute more than $\sim1-2$\% to the total CXB flux from the whole sky. 
Therefore we 
have not excluded them from our analysis.

Measured averaged difference between the large and the small solid angle 
parts of layer 1 MED detector equals to 
 $C_{\rm L}-C_{\rm S}=2.23$ cnts/s. Therefore $C_{\rm CXB, S}=1.75$ cnts/s.
In order to convert this count rate into the physical units (erg
s$^{-1}$cm$^{-2}$ FOV$^{-1}$) we should assume the shape of the CXB spectrum.
The best measurement  of the CXB spectrum in such a broad energy range 
(2--60 keV in our case) was done by Marshall et al. (1980). 
The CXB spectrum was empirically described by a thermal bremsstrahlung model
with the temperature $kT=40$ keV. Below we will always assume this shape
of the CXB spectrum for our analysis.

Using this shape of the CXB spectrum and the response matrix of layer 1 MED
detector we can convert the observed count rate
into the CXB intensity
in the energy range 2--10 keV
$I_{\rm CXB}=8.92\times 10^{-11}$ erg s$^{-1}$cm$^{-2}$ FOV$^{-1}$. 
The effective solid angle of the small
FOV  of MED(layer1) is $4.15$ deg$^2$, that gives the estimation of the CXB intensity from the first layer
of MED $I_{\rm CXB, MED,M1}=2.15\times 10^{-11}$ erg s$^{-1}$cm$^{-2}$ deg$^{-2}$.

The same set of calculations give for the second layer of MED:
$I_{\rm CXB, MED, M2}=1.97\times 10^{-11}$ erg s$^{-1}$cm$^{-2}$ deg$^{-2}$.

The similar calculations for HED detectors give values presented in Table 2.

\begin{table}
\caption{Values of the CXB intensity in the energy band 2-10 keV, obtained from different layer of different detectors of HEAO1/A2}
\tabcolsep=5mm
\begin{tabular}{l|c|c|c}
&MED&HED1&HED3\\
\hline
Layer1&$2.15$&$1.82$&$1.98$\\
\hline
Layer2&$1.97$&$1.90$&$1.91$\\
\end{tabular}

\begin{list}{}
\item -- All values in the units of $10^{-11}$ erg s$^{-1}$cm$^{-2}$ deg$^{-2}$.
\end{list}
\end{table}

\section{Discussion}

Accurate measurements of the CXB intensity are quite complicated. 
Main problems can be divided into three parts:
1)  subtraction of the internal instrumental background, 2)
accurate measurement of the effective solid angle of the instrument 
(including so called stray light contribution to the count rate 
detected by X-ray telescopes) and 3) accurate measurement of the 
instrument effective area and its dependence on energy.

In our approach we accurately determined 
solid angles of the detectors using the celestial calibration source 
(Crab nebula) and accurately subtracted the instrumental 
background (because of the special design of the instrument). 
Stating that the Crab nebula spectrum have the adopted shape 
($dN/dE= 10 E^{-2.05}$ phot s$^{-1}$ cm$^{-2}$ keV$^{-1}$)
in the energy range 2-60 keV
the uncertainties of the obtained effective area values depend
only the knowledge of the detectors response functions.

We have 6 independent measurements of the
CXB using MED, HED1 and HED3 detectors and can try to estimate total
uncertainty of the obtained CXB intensity by calculating the rms deviation of the individual 
measurements from the average one. We obtain CXB intensity 
$I_{\rm CXB}=1.96\pm0.10$ 
erg s$^{-1}$cm$^{-2}$ deg$^{-2}$.

\subsection{Comparison with collimated experiments}

First generations of X-ray instruments, which was represented mainly by 
collimated 
spectrometers could overcome problem number (2) and (3). Comparison of the 
Crab nebula count rates, which spectrum and the flux is considered stable, 
can provide us an accurate crosscalibration if we know the response function 
of the instrument (even without accurate knowledge of the instrument 
effective area). The effective solid angle of the collimator also can 
be measured directly from observations and compared to that of other 
instruments.

In the view of such simplification it is interesting to compare 
the measurements of the CXB intensity made by collimated spectrometers.

We combined the values of CXB intensity measured by these experiments 
(rockets -- Gorenstein et al. 1969, Palmieri et al. 1971, McCammon et al. 1983,
HEAO1 -- this work, RXTE/PCA -- Revnivtsev et al. 2003)
in Table 3 along with their claimed Crab nebula fluxes. The flux
of the Crab nebula presented in these works 
can be used in order to recalculate the effective areas
of the instruments. In order 
to obtain corrected CXB intensity value we divided the CXB intensity 
value provided by authors by the ratio of their
Crab nebula flux in 2-10 keV energy band to $2.39\times 10^{-8}$ 
erg s$^{-1}$ cm$^{-2}$, which we adopted here.
The obtained corrected values of the CXB intensity (Table 3) are remarkably
consistent with each other. The deviation of the values
does not exceed 2$\sigma$ level. It is important to note here 
that uncertainties
of the effective solid angles of rocket flights experiments might
lead to another correction which is not possible for us to make here.

 Relatively strong difference between the CXB intensity value 
presented in original work on 
HEAO1/A2 data ($I_{\rm CXB}\sim (1.67\pm0.17) \times 10^{-11}$ ergs s$^{-1}$cm$^{-2}$ deg$^{-2}$) and the result presented in this paper 
is most likely caused by different assumptions on the 
normalization of the Crab nebula spectrum. Unfortunately  in the original
work of 
Marshall et al. (1980) information about this normalization factor is absent.

\subsection{Comparison with focusing telescopes}

During last decades there were done a number of measurements of the 
CXB intensity with the help of focusing telescopes. 

In the Table 3 we summarize the CXB intensity values obtained
by ASCA, BeppoSAX and XMM-Newton observatories.

Crosscalibration of effective areas of ASCA and BeppoSAX instruments
with those of the collimated experiments can be done with the 
help of Crab nebula. We should remember that after
this cross check there are still uncertainties
in the effective solid angles of focusing telescopes, which can not
be overcome by comparison of the Crab nebula fluxes.

 Crosscalibration of the XMM-Newton instruments with the collimated
spectrometers and with ASCA and BeppoSAX is less clear, because
it can not be done via measurements of the Crab nebula. 
However, for such purpose one can 
use strictly simultaneous observations of weak pointlike objects.
For example we can use simultaneous XMM and RXTE observations of quasar 
3C273 (\cite{tlc03}). The rescaling factor, determined in the paper
of Courvoisier et al. (2003) can not be used by us here because the
flux of the Crab nebula assumed by the authors (which is build-in the
LHEASOFT 5.2 package tasks used in that paper) 
was higher than the value assumed by us (see discussion of this topic 
in \cite{mikej03}). Our estimate
of the RXTE/PCA-XMM/EPIC\_PN rescaling factor 
 is $1.17\pm0.10$  -- the PCA flux of a source 
is $17\pm10$\% higher than that of EPIC-PN. 

\begin{table}
\caption{Summary of measurements of the CXB intensity with different observatories.
CXB$_{\rm corr}$ is the CXB intensity corrected for rescaling factors between observatories (see text)}
\tabcolsep=2mm
\begin{tabular}{l|c|c|c|c}
Instrument&Ref.&CXB&Crab&CXB$_{\rm corr}$\\
\hline
Rockets&1&$1.57\pm0.15$&2.27&$1.65\pm0.16$\\
Rockets&2&$1.90\pm0.20$& $-$&$-$\\
Rockets&3&$2.20\pm0.20$&2.29&$2.29\pm0.21$\\
HEAO/A2&4&$1.96\pm0.10$&2.39&$1.96\pm0.10$\\
RXTE/PCA&5&$1.94\pm0.19$&2.39&$1.94\pm0.19$\\
ASCA/GIS&6&$1.94\pm0.19$&2.16&$2.14\pm0.21$\\
BeppoSAX&7&$2.35\pm0.12$&2.01&$2.79\pm0.14$\\
XMM/EPIC-PN&8&$2.15\pm0.16$&$-^a$&$2.51\pm0.30$\\
XMM/EPIC-PN&9&$2.24\pm0.16$&$-^a$&$2.62\pm0.18$\\
\hline
\end{tabular}
\begin{list}{}
\item - CXB intensity is expressed in units $10^{-11}$ erg s$^{-1}$cm$^{-2}$ deg$^{-2}$, the Crab nebula flux is is units $10^{-8}$ erg s$^{-1}$cm$^{-2}$. Energy band 2-10 keV
\item References: 1 -- Gorenstein et al. 1969, 2 -- Palmieri et al. 1971, 3 -- 
\cite{mccammon83}, 4 -- this work, 5 -- Revnivtsev et al. 2003, 6 -- 
Kushino et al. 2002, 7 -- \cite{vecchi99}, 8 --  Lumb et al. 2002, 9 -- 
De Luca \& Molendi 2004
\item $^a$ - rescaling was done using observations of 3C273, see text
\end{list}
\end{table}

Considering the abovementioned 
numbers with quoted uncertainties we can conclude that there are 
some indications that the intensity of the CXB measured by focusing
telescopes is higher than that measured by collimated experiments.
Such discrepancy is now limited to $\sim10-15$\% and practically
does not exceed $2\sigma$ confidence limits of individual measurements.
The nature of this discrepancy is still unknown. One of the possible
reasons can be the extreme complexity of measurement of 
effective solid angles (stray light effects) of focusing telescopes in comparison with those
of collimated experiments.

\begin{acknowledgements}
M.R. thanks Pavel Shtykovskiy for his help with XMM data analysis.
We thank anonimous referee who help us to significantly improve the paper.
This research has made use of data obtained through the High Energy
Astrophysics Science Archive Research Center Online Service,
provided by the NASA/Goddard Space Flight Center.
\end{acknowledgements}

\end{document}